\documentclass[reprint,runinaddress,nofootinbib,amsmath,amssymb,aps,pra]{revtex4-1}
\usepackage{graphicx}
\usepackage{amsmath}
\usepackage{mathptmx}
\usepackage{array}
\usepackage{amssymb}
\usepackage{color}
\usepackage{float}
\usepackage{dcolumn}
\usepackage{bm}

\usepackage{enumitem} 
\usepackage{tabularx}
\usepackage{soul}
\usepackage[dvipsnames]{xcolor}

\begin{document}

\title{Remote laser-speckle sensing of heart sounds for health assessment and biometric identification.}
\author{Lucrezia Cester$^{1}$, Ilya Starshynov$^{1}$, Yola Jones$^{2}$, Pierpaolo Pellicori$^{2}$, John GF Cleland$^{2}$, Daniele Faccio$^{1}$}
\email{Daniele.Faccio@glasgow.ac.uk}
\affiliation{ $^{1}$ School of Physics and Astronomy, University of Glasgow, Glasgow, UK\\
$^{2}$Robertson Centre for Biostatistics and Clinical Trials, University of Glasgow, Glasgow, UK}
\date{\today}

\begin{abstract}
Assessment of heart sounds is the cornerstone of cardiac examination, but it requires a stethoscope, skills and experience, and a direct contact with the patient. We developed a contactless, machine-learning assisted method for heart-sound identification and quantification based on the remote measurement of the reflected laser speckle from the neck skin surface in healthy individuals. We compare the performance of this method to standard digital stethoscope recordings on an example task of heart-beat sound biometric identification. We show that our method outperforms the stethoscope even allowing identification on the test data taken on different days. This method might allow development of devices for remote monitoring of cardiovascular health in different settings.   
 \end{abstract}
\maketitle
 
\section{Introduction}
Cardiovascular diseases (CVDs) are the leading cause of disability and death worldwide, taking an estimate of 17.9 million lives each year \cite{pagidipati2013estimating,writing2012our}.  Early identification of pathological cardiac conditions might improve well-being and prevent premature deaths. 
The evaluation of heart sounds is the cornerstone of cardiac examination. Normal heart sounds are low-frequency transient mechanical vibrations generated by the closure of heart valves, and should be distinguished from heart murmurs, typically higher frequency, noise-like sounds that are caused by turbulent blood flow~\cite{heart1,Ergen}. The auscultation of heart sounds with a stethoscope is considered to be a clinical `art' with considerable training required in order to distinguish normal from pathological heart sounds and murmurs~\cite{stethoscope1}. Digital stethoscopes and phonocardiographs have also been developed to provide reliable graphical representations of heart sounds and heart sound diagnostics~\cite{666,brusco2004digital,reed2004heart,potes2016ensemble,son2018classification}.  \\
\begin{figure*}[t]
\centering\includegraphics[width=\textwidth]{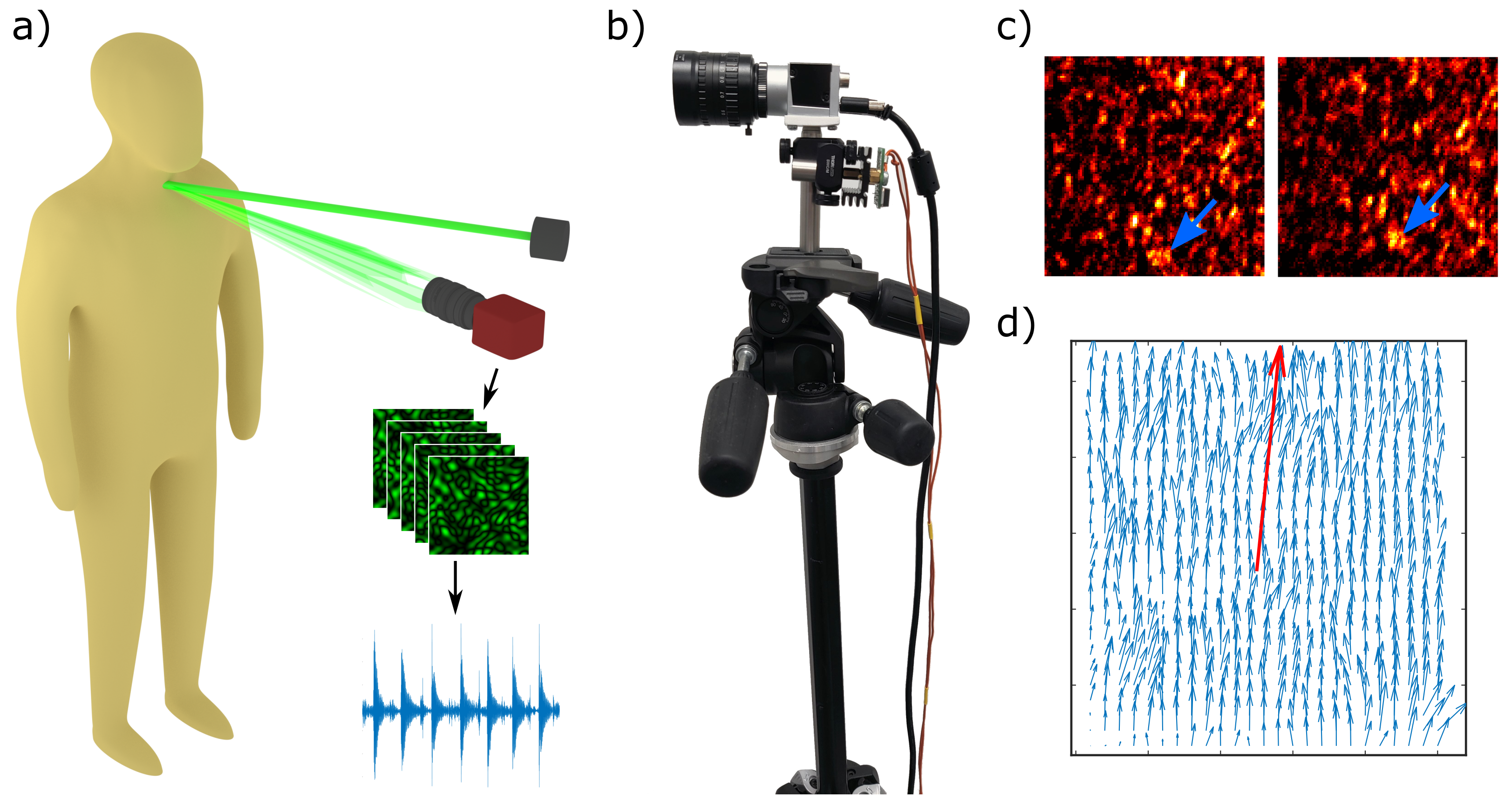}
\caption{(a) and (b) show the experimental set up: a CMOS camera records the speckle pattern reflected from a person's neck and created by a 4 mW continuous 532 nm laser mounted next to the camera (b). The displacement of the skin surface due to the heart sound causes proportional displacements of the speckle pattern at the far field due to speckle memory effect~\cite{freund_memory_1988}. The speckle displacement can thus be be tracked to recover the heart sound.  (c) shows two consecutive frames of the raw data speckle image recording and (d) shows a map of the local displacement calculated using Farneback algorithm. Red arrow shows the average displacement. \label{fig1}}
\end{figure*}
In more detail, the first heart sound (S1 - with a tone that can range between 10 and 140 Hz) originates from the closure of the mitral and tricuspid valves and is separated by the systolic pause from the second heart sound (S2 - with a tone that can range between 10 and 400 Hz), caused by the closure of the aortic and pulmonary valves~\cite{8994674}. Some people can have a third heart sound (S3), which can be either normal or sign of a disease, while additional sounds (ie: S4) and high frequency murmurs (can range between 20 to 1000 Hz~\cite{1139743615}), if identified, often indicate cardiac pathology~\cite{conn2009cardiac}.\\
A heart sound is best appreciated with a stethoscope close to its point of origin, however vibrations produced inside the heart, particularly those of low frequency, can still propagate peripherally through the arteries from where they could be measured. { Most of the existing methods of remote detection of those sounds based on radar~\cite{will2018radar} or visible light~\cite{7868713,333,444,10.1117/12.2253407} acquire only the lowest frequency vibrations, thus providing information only about the heart rate.}
\cite{will2018radar,7868713,333,444,10.1117/12.2253407}. {On the other hand}, Zalevsky et al. and Bianchi et al. have  proposed two different methods for acquisition of sound from the mechanical movement of objects with visible light capable of capturing sounds in 0-1.2~ kHz~\cite{zalevsky2009simultaneous} and 0-5 kHz~\cite{bianchi2019long,bianchi2014vibration} frequency range. {Both of these approaches rely on the detection of the random speckle pattern that is reflected back to the observer's camera and that is generated by random multipath interference from the object (e.g. scattering from the skin). The two approaches differ in how they track the changes in the speckle pattern due to subtle mechanical vibrations of the skin by e.g. using cross-correlation between different images or tracking the centre of mass of a single large speckle. Zalevksy et al. have also reported on the use of laser speckle reflected from the wrist to measure the heart rate ~\cite{10.1117/12.2253407,zalevsky2009simultaneous}. }\\
In this work we assess the feasibility of using these remote detection methods to monitor heart sounds for diagnostic purposes. As the heart sound signal is quite complex it's hard to quantify the recording quality using conventional criteria such as signal to noise ratio or mean square error. We therefore compare the ability of a machine learning algorithm to perform biometric identification~\cite{abo2014biometric,phua2008heart,el2010hsas,gautam2013biometric}, using the laser-speckle detection method and a digital stethoscope recording dataset (HSCT-11)~\cite{spadaccini2013performance}.\\
Here we report on heart sound information that is acquired contactlessly from a distance of around 1 m by shining a weak laser beam at the frontal region of the subject's neck and recording the back-reflected speckle pattern with a high frame-rate CMOS camera. A gradient-based technique is applied to track the `flow' of the speckle pattern over time~\cite{flow}, which contains data on heart sound. For the biometric identification algorithm we use wavelet scattering transform feature extraction, paired with a Support Vector Machine (SVM) classifier. The comparison of this algorithm applied to our data and the standard stethoscope dataset shows that our laser-speckle detection method outperforms the latter.   \\ 
\section{Experimental setup}
The experimental setup is shown in Fig.~\ref{fig1}(a) and Fig.~\ref{fig1}(b) shows a photograph of the actual laser/camera system. A laser diode (DJ532-40 Thorlabs) is directed at the neck of the participant creating an illumination spot of $\sim5$ mm diameter. A camera (Basler acA640-750um, Germany) collects the resulting dynamic speckle pattern at $f_{\text{samp}}$ = 1.5 kHz frame rate with 200 $\times$ 208 pixel resolution. The acquisition frame-rate is chosen to be as high possible whilst still delivering a  good signal-to-noise ratio images from the low-power illumination laser (limited to 4 mW). A standard objective with focal length $f=25$ mm, 0.95 f-stop allows to detect the reflected speckle field from 10 cm up to 5 m away from the subject (larger distances were not tested but it has been shown by  Bianchi et al. and  Zalevsky et al. that the reflected speckles can be detected at distances up to 300 m~\cite{bianchi2019long,bianchi2014vibration,zalevsky2009simultaneous}). In the current experiment the camera was located at around 1 m distance from the test subject and collected the light at around 20 cm defocus distance from the illumination spot. The test subject is seated in a chair in a natural pose while the device records the time dynamics of the speckle pattern from their skin. 
The resulting speckle recordings are then post-processed to retrieve the heart sounds. Two example successive frames as captured on the camera are shown in Fig.~\ref{fig1}(c): the blue arrow indicates an example feature in the speckle pattern that highlights the shift from one frame to the next. \\
\section{Data processing}
\subsection{Retrieval of the raw sound signal from the speckle frame sequence}
\begin{figure}[t!]
\centering
\includegraphics[width=0.45\textwidth]{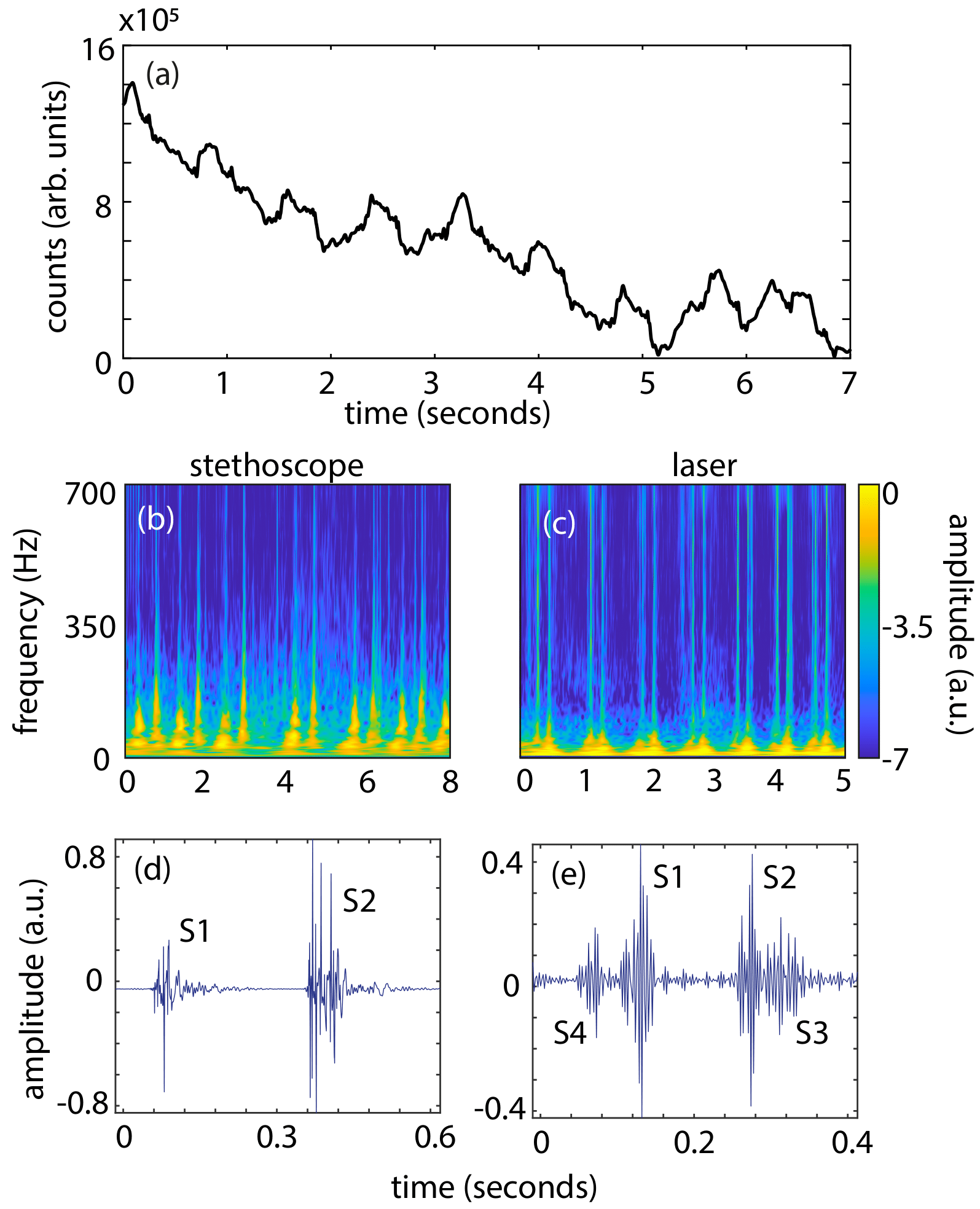}
\caption{Heart sound recordings obtained with a stethoscope (left) and laser (right). (a) shows the raw-data speckle displacement over time, thus resulting from both heart beat sound and macroscopic human body movements. (b) and (c) show respectively the scalograms in log scale of the sound acquired with a stethoscope from the subject's chest and with our device from the subject's neck. (d) and (e) show cropped time traces corresponding to the signals in (b) and (c) respectively.  \label{fig2} }
\end{figure}

The first step of the data processing in our method is the retrieval of the raw sound signal from the recorded speckle frame sequence. Since the scattering surface (the subject's neck) is subject to small deformations due to heart sound pressure waves propagating in the blood vessels, we can rely on the so-called speckle memory effect~\cite{freund_memory_1988}, i.e.  the speckle pattern does not change shape upon tilting or vibration of the skin surface but translates proportionally to the tilting angle~\cite{prunty2017demystifying}. We employed a speckle tracking algorithm based on the optical flow~\cite{flow}, implemented in MATLAB (2018b) Computer Vision Toolbox to retrieve the local displacement map at each pixel of the image and then averaged these vectors to calculate the overall displacement amplitude and direction between consecutive frames, see Fig.~\ref{fig1}(d). An example of the speckle displacement retrieved from the optical flow is shown in Fig.~\ref{fig2}(a), and in Fig~\ref{fig2}(c) we show its scalogram. In Fig.~\ref{fig2}(b) we show for comparison a typical heart sound acquired with a digital stethoscope positioned on the chest \cite{Goldberger}. Comparison of Fig.~\ref{fig2}(b) with Fig.~\ref{fig2}(c) indicates that our method gives comparable signal-to-noise ratio across the main 20-700 Hz region. The optical flow also picked up macroscopic movements of the test subjects, therefore giving a large contribution in the 0-20 Hz range, which we eliminated by passing the signals through a 20-700 Hz band-pass filter (Butterworth, order 10). The resulting heart beat signal is shown in Fig.~\ref{fig2}(e) and for comparison a stethoscope signal is shown in Fig.~\ref{fig2}(d). 
Out of our pool of 10 subjects, 1 subject, whose heart sound recording we show in Fig.~\ref{fig2}(c) and Fig.~\ref{fig2}(e), presented not only S1 and S2, but also S3 and S4 sounds. The additional S3 and S4 signals are very clear and only visible after the filtering process described above. {As a result of this incidental finding the test subject was referred for further clinical investigation and underlines the potential future of this method for identifing pathological heart sounds, murmurs and rhythm abnormalities in patients with a variety of cardiac pathologies}.\\
\subsection{Database acquisition for biometric identification}
 We acquired heart sounds from 10 subjects with the experimental setup shown in Fig.~\ref{fig1}. 
{ For each of the subjects we recorded 4.5 minutes of cardiac activity in one day and 30 sec in the following 1-2 days.}
 In both of these sessions, the laser was pointed towards the base of the subject's neck without attempting to reproduce the precise location of the laser spot between sessions.
{Each of the recordings was bandpass-filtered as described above, rescaled to unit amplitude and cut into 2.5 sec segments thus providing a dataset of 108 recordings per person taken in one day and 12 recordings taken on another day.\\
In order to compare the quality of the data taken with our method to commonly used stethoscope recordings we used an open HSCT-11 dataset~\cite{spadaccini2013performance}, containing digital stethoscope (ThinkLabs Rhythm Digital Electronic Stethoscope) recordings taken from 206 people. We have selected, arbitrarily, data from 10  subjects for whom there was at least 2.5 min of cardiac activity recording from this dataset. We segmented these recordings into 3 second pieces thus obtaining a dataset of 45 recordings per person. These recordings, being in WAVE format were already scaled from -1 to 1 and, as the digital stethoscope has its own filtering algorithm, we did not apply any additional filtering to this data. \\
} 
\begin{figure*}[th!]
\centering
\includegraphics[width=\textwidth]{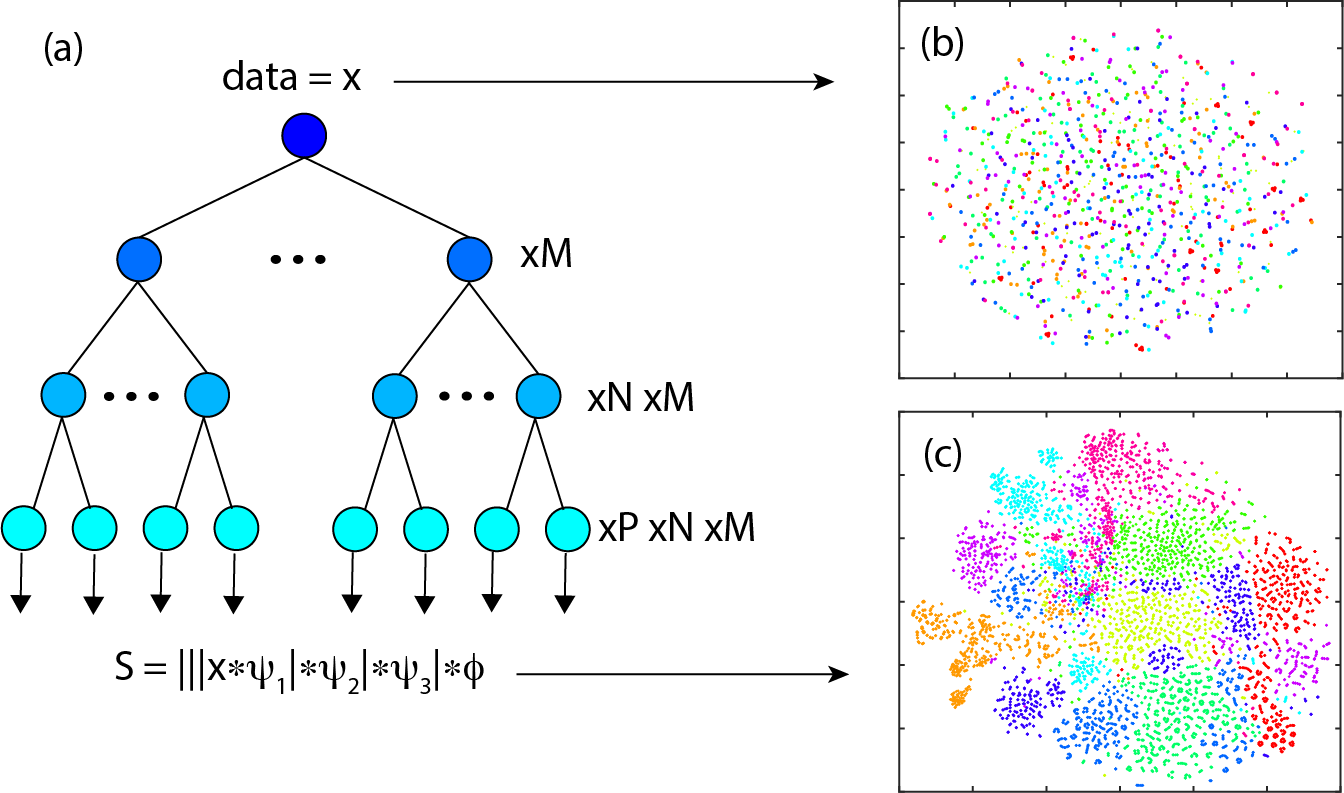}
\caption{ (a) shows on the left the architecture of the scattering transform. The signal is convolved with M=56 filters in the first layer, N=30 in the second and P=9 in the third.  The 2D embedding of the raw heart sound data from multiple individuals obtained using T-SNE algorithm is shown in (b) where each color corresponds to a different person. No clear clustering is observed in that case. However, after going through the scattering transform, {as demonstrated in (c)}, the data within the same class is clustered together and the classification problem is simplified. \label{fig3} }
\end{figure*}
\section{Biometric identification algorithm}
The algorithm we used to identify people contained two major steps: feature extraction using wavelet scattering transform, implemented in MATLAB Wavelet Toolbox, and classification using a support vector machine (SVM), implemented in MATLAB Statistics and Machine Learning Toolbox.\\ 
\subsection{Feature extraction}
Feature extraction identifies stable features and disregards signal deformations due to for example additive noise, translations, dilations, etc. We used a wavelet scattering network to extract features of our PCGs~\cite{bruna2013invariant}. The architecture of our wavelet scattering transform, which resembles the physiological processing method used by the cochlea, uses three layers of wavelet filter banks (Gabor mother wavelet) with M=56, N=30 and P=9 filters/node in each of the layers. The extracted coefficients  allow to group signals belonging to the same class closer together { by means of a dimensionality reduction technique}. This concept is illustrated with an example in Fig.~\ref{fig3}.  In Fig.~\ref{fig3}(a) we show an example of the raw data represented after dimensionality reduction using t-SNE (MATLAB Statistics and Machine Learning Toolbox) \cite{tSNE}. When compared to the same visualisation after passing through the wavelet scattering network shown in Fig.~\ref{fig3}(b), we see a clear change from a disorganised to a strongly organised grouping of the data.\\

 \subsection{Classification Algorithm}
Once features are extracted with the wavelet scattering transform we use a SVM to fit the regions corresponding to different classes within the feature space. We used a third degree polynomial kernel SVM with hinge loss which was found to give optimal results.  \\
\begin{figure*}[t]
\centering
\includegraphics[width=\textwidth]{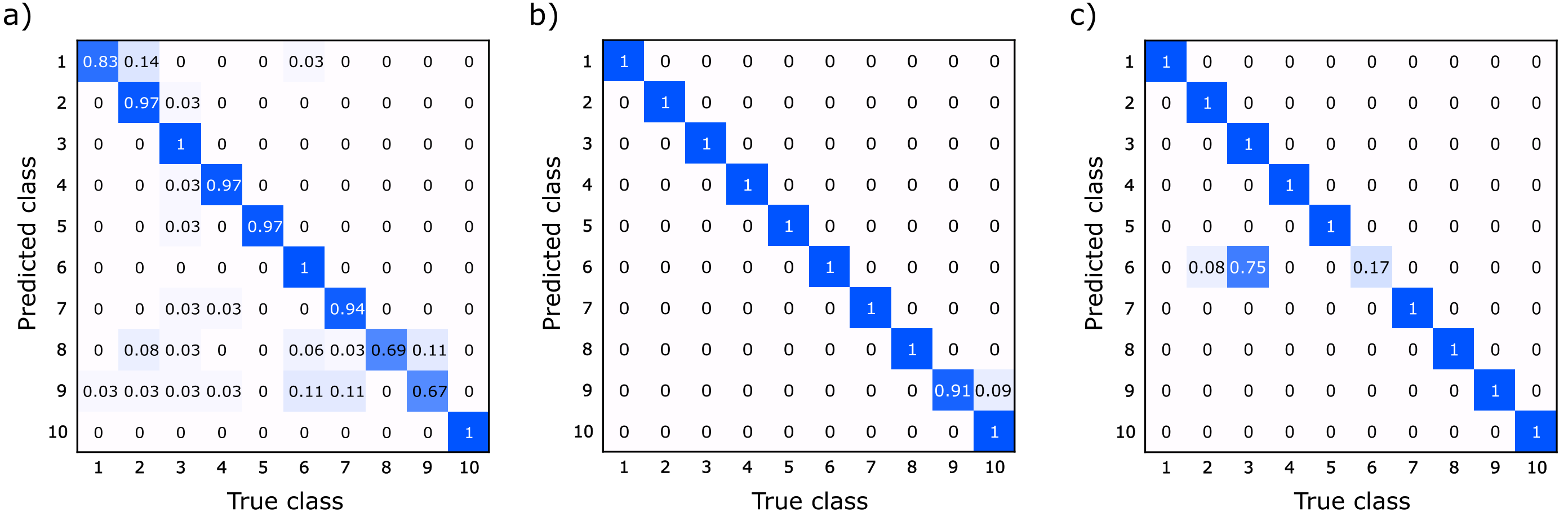}
\caption{Confusion matrices of the biometric identification algorithm for the three testing sets. The algorithm was trained on 30 sec of heart-beat sounds per person for remote and stethoscope methods. (a) shows the CM for 10 arbitrary subjects taken from the HSCT-11 open heart-beat sound dataset, and tested on 35 3 sec recordings (90.6\% accuracy). (b) shows the CM for the remote detection method with the test data (86 2.5 sec recordings) taken on the same date as the training (99.1\% accuracy). (c) same as (b) but for next day testing data (12 2.5 sec recordings, 91.7\% accuracy)  \label{fig4}}
\end{figure*}
\section{Results}
{In Fig.~\ref{fig4} we show the confusion matrices (CM) of the testing data passed through our biometric identification algorithm. We show in (a) the CM of the stethoscope recording dataset, in (b) the CM of our remote detection method tested on the same day recording, in (c) the remote detection method tested on different day recordings. For both methods we trained the algorithm on 30 sec of heart beat recordings (taken in the first day for laser-speckle method). The classification accuracy for remote detection data was 99.1\% (tested on 86 2.5 sec recordings) and 91.7\% (tested on 12 2.5 sec recordings) for the same day and another day training datasets respectively and 90.6\% (tested on 35 3 sec recordings) for the stethoscope data. As can be seen from this figure, our method outperforms digital stethoscope in this task even when the heart sounds are taken on a different day. This indicates the ability of this method to capture fine features of heart sounds and its potential as a tool for monitoring cardiac health. }\\
\section{Conclusions}
Heart sounds are a remarkably complex signature of cardiac health and, when captured in full detail, can provide access to a range of diagnostic opportunities including heart health monitoring and even biometric identification. We have developed a contactless optelectronic sensing approach with a data processing pipeline that allows to extract high quality heart sound signals remotely from the neck area, bypassing the need for precordial, contact based auscultation. We compare the data obtained with our method to the standard stethoscope recordings in the biometric identification task, showing that we can achieve better accuracy even when testing data is taken on a different day for a shorter periods of time with respect to the training data. Future work will look into further exploiting the full potential of these optoelectronic approaches, including identification of pathological heart sounds, murmurs and abnormal heart rhythms. The hope is that in the near future related technologies could become part of the lived environment with a route towards continuous health assessment for precision medicine.\\

\section{Acknowledgments}
DF acknowledges financial support from the Royal Academy of Engineering Chair in Emerging Technologies Scheme.
This work was supported by the Engineering
and Physical Sciences Research Council of the UK (EPSRC) Grant Nos. EP/T021020/1, EP/T00097X/1, EP/S026444/1
and the UK MOD University Defence Research Collaboration (UDRC) in Signal Processing. PP and YJ are supported by an MRC grant in precision medicine (grant number: 2285827).\\
This research was approved by the University of Glasgow ethics approval committee, ethical approval application no.300200122.
\section*{Disclosures}
The authors declare no conflicts of interest
\section*{Data availability}
The data and the source code created for this research are available at \url{http://dx.doi.org/10.5525/gla.researchdata.1238}

\end{document}